**22**

# Mobility and Handoff Management in Wireless Networks

Jaydip Sen
*Tata Consultancy Services*
*INDIA*

## 1. Introduction

With the increasing demands for new data and real-time services, wireless networks should support calls with different traffic characteristics and different Quality of Service (QoS) guarantees. In addition, various wireless technologies and networks exist currently that can satisfy different needs and requirements of mobile users. Since these different wireless networks act as complementary to each other in terms of their capabilities and suitability for different applications, integration of these networks will enable the mobile users to be always connected to the best available access network depending on their requirements. This integration of heterogeneous networks will, however, lead to heterogeneities in access technologies and network protocols. To meet the requirements of mobile users under this heterogeneous environment, a common infrastructure to interconnect multiple access networks will be needed. Although IP has been recognized to be the de facto protocol for next-generation integrated wireless, for inter-operation between different communication protocols, an adaptive protocol stack is also required to be developed that will adapt itself to the different characteristics and properties of the networks (Akyildiz et al., 2004a). Finally, adaptive and intelligent terminal devices and smart base stations (BSs) with multiple air interfaces will enable users to seamlessly switch between different access technologies.

For efficient delivery of services to the mobile users, the next-generation wireless networks require new mechanisms of *mobility management* where the location of every user is proactively determined before the service is delivered. Moreover, for designing an adaptive communication protocol, various existing mobility management schemes are to be seamlessly integrated. In this chapter, the design issues of a number of mobility management schemes have been presented. Each of these schemes utilizes IP-based technologies to enable efficient roaming in heterogeneous network (Chiussi et al., 2002). Efficient handoff mechanisms are essential for ensuring seamless connectivity and uninterrupted service delivery. A number of handoff schemes in a heterogeneous networking environment are also presented in this chapter.

The chapter is organized as follows. Section 2 introduces the concept of mobility management and its two important components- *location management* and *handoff management*. Section 3 presents various network layer protocols for macro-mobility and micro-mobility. Section 4 discusses various link layer protocols for location management.

Section 5 introduces the concept of handoff. Different types of handoff mechanisms are classified, and the delays associated with a handoff procedure are identified. Some important cross-layer handoff mechanisms are also discussed in detail. Section 6 presents *media independent handover* (MIH) services as proposed in IEEE 802.21 standards. It also discusses how MIH services can be utilized for designing seamless mobility protocols in next-generation heterogeneous wireless networks. Section 7 discusses security issues in handover protocols. Section 8 identifies some open areas of research in mobility management. Section 9 concludes the chapter.

## 2. Mobility Management

With the convergence of the Internet and wireless mobile communications and with the rapid growth in the number of mobile subscribers, mobility management emerges as one of the most important and challenging problems for wireless mobile communication over the Internet. Mobility management enables the serving networks to locate a mobile subscriber's point of attachment for delivering data packets (i.e. location management), and maintain a mobile subscriber's connection as it continues to change its point of attachment (i.e. handoff management). The issues and functionalities of these activities are discussed in this section.

### 2.1 Location management

Location management enables the networks to track the locations of mobile nodes. Location management has two major sub-tasks: (i) *location registration*, and (ii) *call delivery* or *paging*. In location registration procedure, the mobile node periodically sends specific signals to inform the network of its current location so that the location database is kept updated. The call delivery procedure is invoked after the completion of the location registration. Based on the information that has been registered in the network during the location registration, the call delivery procedure queries the network about the exact location of the mobile device so that a call may be delivered successfully. The design of a location management scheme must address the following issues: (i) minimization of signaling overhead and latency in the service delivery, (ii) meeting the guaranteed quality of service (QoS) of applications, and (iii) in a fully overlapping area where several wireless networks co-exist, an efficient and robust algorithm must be designed so as to select the network through which a mobile device should perform registration, deciding on where and how frequently the location information should be stored, and how to determine the exact location of a mobile device within a specific time frame.

### 2.2 Handoff management

Handoff management is the process by which a mobile node keeps its connection active when it moves from one access point to another. There are three stages in a handoff process. First, the initiation of handoff is triggered by either the mobile device, or a network agent, or the changing network conditions. The second stage is for a new connection generation, where the network must find new resources for the handoff connection and perform any additional routing operations. Finally, data-flow control needs to maintain the delivery of the data from the old connection path to the new connection path according to the agreed-upon QoS guarantees. Depending on the movement of the mobile device, it may undergo various types of handoff. In a broad sense, handoffs may be of two types: (i) intra-system handoff (horizontal handoff) and (ii) inter-system handoff (vertical handoff). Handoffs in homogeneous networks are referred to as intra-system handoffs. This type of handoff occurs when the signal strength of the serving BS goes below a certain threshold value. An inter-system handoff between heterogeneous networks may arise in the following scenarios (Mohanty, 2006) - (i) when a user moves out of the serving network and enters an overlying network, (ii) when a user connected to a network chooses to handoff to an underlying or overlaid network for his/her service requirements, (iii) when the overall load on the network is required to be distributed among different systems.

The design of handoff management techniques in all-IP based next-generation wireless networks must address the following issues: (i) signaling overhead and power requirement for processing handoff messages should be minimized, (ii) QoS guarantees must be made, (iii) network resources should be efficiently used, and (iv) the handoff mechanism should be scalable, reliable and robust.

**2.3 Mobility management at different layers**

A number of mobility management mechanisms in homogeneous networks have been presented and discussed in (Akyildiz et al., 1999). Mobility management in heterogeneous networks is a much more complex issue and usually involves different layers of the TCP/IP protocol stack. Several mobility management protocols have been proposed in the literature for next-generation all-IP wireless networks. Depending on the layers of communication protocol they primarily use, these mechanisms can be classified into three categories – protocols at the networks layer, protocols at the link layer and the cross-layer protocols. Network layer mobility protocols use messages at the IP layer, and are agnostic of the underlying wireless access technologies (Misra et al., 2002). Link layer mobility mechanisms provide mobility-related features in the underlying radio systems. Additional gateways are usually required to be deployed to handle the inter-operating issues when roaming across heterogeneous access networks. In link layer protocols, handoff signals are transmitted through wireless links, and therefore, these protocols are tightly-coupled with specific wireless technologies. Mobility supported at the link layer is also called *access mobility* or *link layer mobility* (Chiussi et al., 2002). The cross-layer protocols are more common for handoff management. These protocols aim to achieve network layer handoff with the help of communication and signaling from the link layer. By receiving and analyzing, in advance, the signal strength reports and the information regarding the direction of movement of the mobile node from the link layer, the system gets ready for a network layer handoff so that packet loss is minimized and latency is reduced.

**3. Network Layer Mobility Management Mechanisms**

Over the past several years, a number of IP mobility management protocols have been proposed. Different mobility management frameworks can be broadly distinguished into two categories - device mobility management protocol for localized or *micro-mobility* and protocols for inter-domain or *macro mobility*. The movement of a mobile node (MN) between two subnets within one domain is referred to as micro-mobility. For example, the movement of MN from subnet *B* to subnet *C* in Figure 1 is an example of micro-mobility. An example of micro-mobility in UMTS Terrestrial Radio Access Networks (UTRAN) is movement of an MN from one BS to another, both BSs belonging to the same *random access network* (RAN), while in WLAN it is a node movement between two *access points* (APs). The movement of devices between two network domains is referred to as macro-mobility. For example, the movement of MN from domain 1 to domain 2 in Figure 1 is an example of macro-mobility. A domain represents an administrative body, which may include different access networks, such as WLAN, second-generation (2G), and third-generation (3G) networks (Akyildiz et al., 2004b). Next-generation all-IP wireless network will include various heterogeneous networks, each of them using possibly different access technologies. Therefore, satisfactory macro-mobility solution supporting all these technologies is needed.

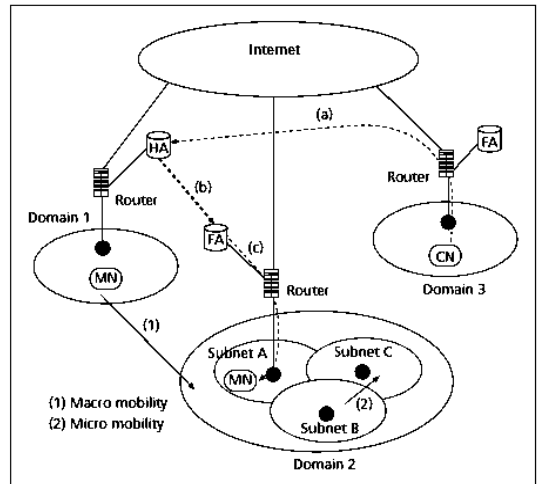

Fig. 1. Mobile IP Architecture [Source: (Akyildiz et al., 2004)]

**3.1 Macro-mobility protocols**
Mobile IP is the most widely used protocol for macro-mobility management. In addition to Mobile IP, three macro-mobility architectures are discussed in the section. These protocols are: Session Initiation Protocol (SIP)-based mobility management, multi-tier hybrid SIP and Mobile IP protocol, and network inter-working agent-based mobility protocol.
**Mobile IP:** Mobile IP (Perkins, 2008) is the most well-known macro mobility scheme that solves the problem of node mobility by redirecting the packets for the MN to its current location. It introduces seven elements: (i) *Mobile node* (MN) – a device or a router that can change its point of attachment to the Internet, (ii) *Correspondent node* (CN) – the partner with which MN communicates, (iii) *Home network* (HN) – the subnet to which MN belongs, (iv) *Foreign network* (FN) – the current subnet in which the MN is visiting, (v) *Home agent* (HA) – provides services for the MN and is located in the HN, (vi) *Foreign agent* (FA) – provides services to the MN while it visits in the FN, (vii) *Care-of-address* (CoA) – defines the current location of the MN; all packets sent to the MN are delivered to the CoA. Mobile IP protocol has three steps: (i) agent discovery, (ii) registration, and (iii) routing and tunneling.
*Agent discovery*: An MN is able to detect whether it has moved into a new subnet by two methods – agent advertisement and agent solicitation. In the agent advertisement method, FAs and HAs advertise their presence periodically using agent advertisement messages.

These advertisement messages can be seen as beacon broadcasts into the subnets. An MN in a subnet can receive agent advertisements. If no agent advertisement messages are found or the inter-arrival time is too high, the MN may send agent solicitations. After the step of agent advertisement or solicitation, the MN receives a CoA. The CoA may be either an FA or a co-located CoA (Perkins, 2008). A co-located CoA is found by using Dynamic Host Configuration Protocol (DHCP) or Point-to-Point Protocol (PPP).

*Registration*: After the MN receives its CoA, it registers it with the HA. The main objective of the registration is to inform the HA about the current location of MN. The registration may be done in two ways depending on the location of the CoA. If the CoA is the FA, the MN sends its registration request to the FA which in turn forwards it to the HA. If the CoA is co-located, the MN may send the request directly to the HA.

*Routing and tunneling*: When a CN sends an IP packet to the MN, the packet is intercepted by the HA. The HA encapsulates the packet and tunnels it to the MN's CoA. With FA CoA, the encapsulated packet reaches the FA serving the MN. The FA decapsulates the packet and forwards it to the MN. With co-located CoA, the encapsulated packets reach the MN, which decapsulates them. In Figure 1, the tunneling (step b) ends at the MN instead of at the FA.

*Paging Extension for Mobile IP*: For saving battery power at MNs, IP paging mechanism has been proposed (Haverinen & Malinen, 2000). Paging typically includes transmitting a request for an MN to a set of locations, in one of which the MN is expected to be present. The set of locations is called a paging area and it consists of a set of neighboring base stations. A network that supports paging allows the MNs to operate in two different states – an active state and a standby state. In an active state, the MN is tracked at the finest granularity such as its current base station (resulting in no need for paging). In the standby state, the MN is tracked at a much coarser granularity such as a paging area. The MN updates the network less frequently in stand by mode (every paging area change) than in active state (every base station). The cost of paging, however, is the complexity of the algorithms and the protocols required to implement the procedures, and the delay incurred for locating an MN.

*Drawbacks of Mobile IP*: The Mobile IP has the following shortcomings:
- The packets sent from a CN to an MN are received by the HA before being tunneled to the MN. However, packets from the MN are sent directly to the CN. This inefficient mechanism of non-optimized Mobile IP is called *triangular routing*. It results in longer routes and more delay in packet delivery.
- When an MN moves across two different subnets, the new CoA cannot inform the old CoA about MN's current location. Packets tunneled to the old CoA are lost.
- Mobile IP is not an efficient mechanism in a highly mobile scenario as it requires an MN to send a location update to the HA whenever it changes its subnet. The signaling cost for location updates and the associated delay may be very high if the distance between the visited network and the home network is large.

*Optimization in Mobile IP*: In (Perkins & Johnson, 2001), an optimization technique has been proposed to solve the problem of triangular routing. The idea is to inform the CN about the current location of the MN so as to bypass the HA. The CN can learn the location of the CoAs of the MN by caching them in a binding cache in the CN. When a CN sends packets to an MN, it first checks if it has a binding cache entry for the MN. If there is an entry, the CN tunnels the packets directly to the CoA. If no binding cache entry is available, the CN sends the packets to the HA, which in turn tunnels them to the CoA. In optimized Mobile IP, the packets tunneled by the HA to the old CoA are not lost in transit. When an MN registers with a new FA, it requests the new FA to notify the previous FA about its movement. As the old FA now knows the location of the current FA, it can forward the packets to the new FA.

**SIP-Based Mobility Management:** In (Salsano et al., 2008), a *Session Initiation Protocol* (SIP)-based solution, called *mobility management using SIP extension* (MMUSE), has been proposed that supports vertical handoffs in next-generation wireless networks. SIP has been chosen by the Third Generation Partnership Project (3GPP) as the signaling protocol to set up and control real-time multimedia sessions. In MMUSE, a *mobile host* (MH) is assumed to be equipped with multiple network interfaces; each of them is assigned a separate IP address when connected to different *access networks* (ANs). The MH uses the SIP protocol to set up multimedia sessions. The architecture of the scheme is depicted in Figure 2. The *session border controller* (SBC) is a device that is typically located at the border of an IP network, and manages all the sessions for that network. A new entity, called the *mobility management server* (MMS) resides within the SBC. The MMS cooperates with another entity – *mobility management client* (MMC) that resides in each MH. Both the SIP *user agents* (UAs) on the MH and on the *corresponding host* (CH) remain unaware of all the handoff procedures, which are handled by the MMC and the MMS. On the MH, the UA sees only the MMC as its outbound proxy and forwards the normal SIP signaling and media flows to it. MMC relays the packets to the MMS/SBC. From there on, the packets follow the path determined by the usual SIP routing procedure. Every time the MH moves across two ANs, a location update SIP message is sent to the MMS. This is done over the new network so that the procedure can be completed even if the old network is suddenly not available. If the MMS receives a call addressed to one of its served MHs, it forwards the call to the correct interface. When the MH changes its AN while it is engaged in a call, the procedure is almost identical. However, in this case, the MMC sends to the MMS an SIP message that contains the additional information required to identify the call to be shifted to new interface. To minimize the handoff duration, the *real-time transport protocol* (RTP) flow coming from the MH during the handoff is duplicated using the MMC. When the MMC starts the handoff procedures, it sends the handover request to the MMS and at the same time, it starts duplicating the RTP packets over both interfaces. As soon as the MMS receives the handover message, the packets coming from the new interface are already available. The MMS performs the switching and sends the reply back to the MMC. When the MMC receives the reply message, it stops duplicating the packets.

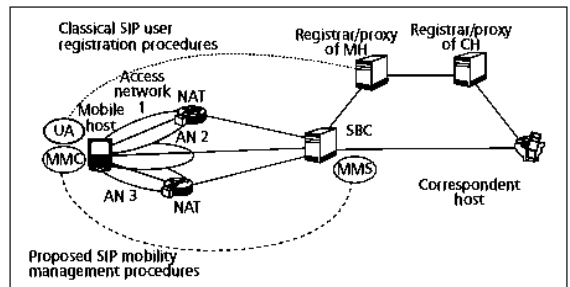

Fig.2. Architecture of MMUSE [Source: (Salsano et al., 2008)]

**Multi-Layer Mobility Management using Hybrid SIP and Mobile IP:** In (Politis et al., 2003), two mobility management architectures based on SIP and Mobile IP are presented.

The two approaches provide mobility in two different layers: application and network layers respectively. The scheme is therefore called multi-layer mobility management scheme. The SIP-based protocol uses SIP in combination with IP encapsulation mechanisms on CHs to support mobility for all types of traffic from/to the MH. The second approach performs separation of traffic and employs SIP in combination with *network address translation* (NAT) mechanisms to support mobility for real-time traffic over UDP. The mobility for non-real-time traffic (mainly TCP-based applications) is supported by Mobile IP. In the SIP-based approach, if the MH moves during a session, the SIP UA sends a SIP re-INVITE request message to each of its CHs. If a CH runs a TCP session, IP encapsulation is used to forward packets to MH. However, if a CH runs a UDP session, the packets are sent directly to the MH's new address. The MH completes the handoff by sending a SIP REGSITER message to the SIP server. For the hybrid SIP/Mobile IP scheme, the inter-domain mobility is based on the synergy of SIP with Mobile IP. Traffic from/to an MH is separated on the domain edge routers. SIP signaling is used to support inter-domain mobility for real-time (RTP over UDP) traffic, while Mobile IP supports non-real-time traffic.

**Network Inter-Working Agent-Based Mobility Management:** In (Akyildiz et al., 2005) an architecture has been proposed for next-generation all-IP wireless systems. Different wireless networks are integrated through an entity called the *network inter-working agent* (NIA). In Figure 3, an NIA integrates one WLAN, one cellular network, and one satellite network. NIA also handles authentication, billing, and mobility management issues during inter-system (inter-domain) roaming. Two types of movement of an MH are considered: movement between different subnets of one domain (intra-domain mobility) and movement between different access networks belonging to different domains (inter-domain mobility). For inter-domain mobility, a novel cross-layer mobility management protocol is proposed, which makes an early detection of the possibility of an inter-domain handoff and allows authentication, authorization and registration of the MH in the new domain before the actual handoff. These interoperability operations are executed by the NIA.

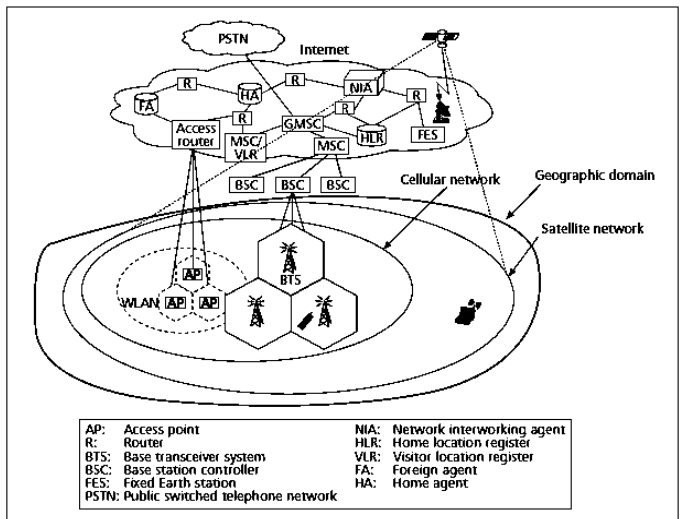

Fig. 3. NIA-Based Mobility Management Architecture [Source: (Akyildiz et al., 2005)]

**3.2 Micro-mobility protocols**
Over the past several years a number of IP micro-mobility protocols have been proposed, designed and implemented that complement the base Mobile IP (Campbell & Gomez, 2001) by providing fast, seamless and local handoff control. IP micro-mobility protocols are designed for environments where MHs changes their point of attachment to the network so frequently that the base Mobile IP mechanism introduces significant network overhead in terms of increased delay, packet loss and signaling. For example, many real-time wireless applications, e.g. VOIP, would experience noticeable degradation of service with frequent handoff. Establishment of new tunnels can introduce additional delays in the handoff process, causing packet loss and delayed delivery of data to applications. This delay is inherent in the round-trip incurred by the Mobile IP as the registration request is sent to the HA and the response sent back to the FA. Route optimization (Perkins & Johnson, 2001) can improve service quality but it cannot eliminate poor performance when an MH moves while communicating with a distant CH. Micro-mobility protocols aim to handle local movement (e.g., within a domain) of MHs without interaction with the Mobile IP-enabled Internet. This reduces delay and packet loss during handoff and eliminates registration between MHs and possibly distant HAs when MHs remain inside their local coverage areas. Eliminating registration in this manner also reduces the signaling load experienced by the network.
The micro-mobility management schemes can be broadly divided into two groups: (i) tunnel-based schemes and (ii) routing-based schemes. In tunnel-based approaches, the location database is maintained in a distributed form by a set of FAs in the access network. Each FA reads the incoming packet's original destination address and searches its visitor list for a corresponding entry. If an entry exists, it is the address of next lower level FA. The sequence of visitor list entries corresponding to a particular MH constitutes the MH's location information and determines the route taken by downlink packets. Mobile IP regional registration (MIP-RR) (Fogelstroem et al., 2006), hierarchical Mobile IP (HMIP) (Soliman et al., 2008), and intra-domain mobility management protocol (IDMP) (Misra et al., 2002) are tunnel-based micro-mobility protocol.
Routing-based approaches forward packets to an MH's point of attachment using mobile-specific routes. These schemes introduce implicit (snooping data) or explicit signaling to update mobile-specific routes. In the case of Cellular IP, MHs attached to an access network use the IP address of the gateway as their Mobile IP CoA. The gateway decapsulates packets and forwards them to a BS. Inside the access network, MHs are identified by their home address and data packets are routed using mobile-specific routing without tunneling. Cellular IP (CIP) (Campbell et al., 2000) and handoff-aware wireless access Internet infrastructure (HAWAII) (Ramjee et al., 2002) are routing-based micro-mobility protocols.
**Mobile IP Regional Registration:** In Mobile IP, an MN registers with its HA each time it changes its CoA. If the distance between the visited network and the home network of the MN is large, the signaling delay for these registrations may be long. MIP-RR (Fogelstroem et al., 2006) attempts to minimize the number of signaling messages to the home network and reduce the signaling delay by performing registrations locally in a regional network. This reduces the load on the home network, and speeds up the process of handover. The scheme introduces a new network node called the *gateway foreign agent* (GFA). The address of the GFA is advertised by the FAs in a visited domain. When an MN first arrives at this visited domain, it performs a home registration - that is, a registration with its HA. At this time, the MN registers the address of the GFA as its CoA. When the MN moves between different

FAs within the same visited domain, it only needs to make a regional registration to the GFA. When the MN moves from one regional network to another, it performs a home registration with its HA. The packets for the MN are first intercepted by its HA, which tunnels them to the registered GFA. The GFA checks its visitor list and forwards the packets to the corresponding FA of the MN. The FA further relays the packets to the MN. The use of the GFA avoids any signaling traffic to the HA as long as the MN is within a regional network.

**Hierarchical Mobile IPv6:** The basic idea of hierarchical Mobile IP (Soliman et al., 2008) (HMIP) is the same as that of regional registration scheme. HMIP introduces a new Mobile IP node called the *mobility anchor point* (MAP). An MN is assigned two CoAs - regional CoA (RCoA) and on-link CoA (LCoA). The MN obtains the RCoA from the visited networks. RCoA is an address on the MAP's subnet. The LCoA is the CoA that is based on the prefix advertised by the *access router* (AR). The AR is the default router of the MN and receives all outbound traffic from it. When an MN enters a new network, it receives router advertisement that contains the available MAPs and their distances from the MN. The MN selects a MAP, gets the RCoA in the MAP's domain and the LCoA from the AR. The MN sends a binding update to the MAP. The MAP records the binding and inserts it in its binding cache (foreign registration). The MAP sends the binding update message also to the MN's HA and to the CNs (home registration). When MN is outside its home network, the incoming data to MN goes through MAP hierarchy. Messages from CN or HA are received by the MAP, which tunnels them to LCoA. As the MN roams locally, it gets a new LCoA from its new AR. The RCoA remains unchanged as long as the MN is within the same network.

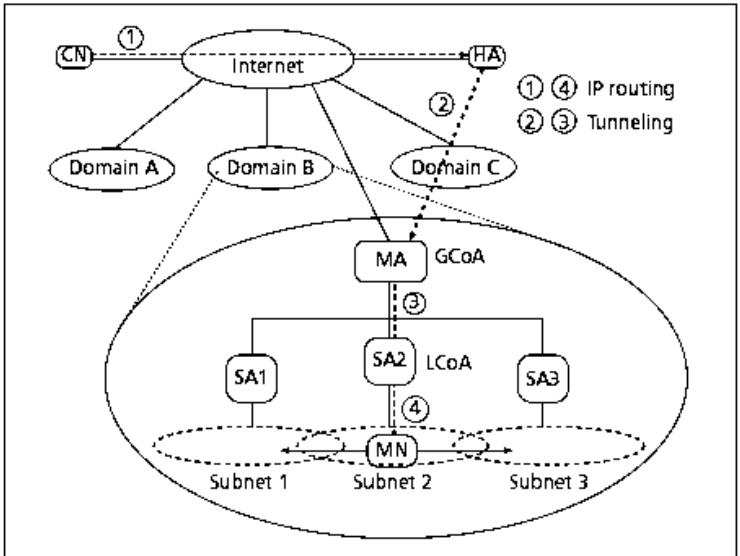

Fig. 4. The Architecture of IDMP [Source: (Akyildiz et al., 2004)]

**Intra-Domain Mobility Management Protocol:** Intra-domain mobility management protocol (IDMP) (Misra et al., 2002) is a two-level, hierarchical, multi-CoA, intra-domain mobility management protocol. The first level of the hierarchy consists of different mobility domains. The second level consists of IP subnets within each domain. This hierarchical approach localizes the scope of intra-domain location update messages and thereby reduces both the global signaling load and update latency. The two-level hierarchical architecture defined by IDMP is shown in Figure 4. IDMP consists of two types of entities: (i) *mobility agent* (MA) and (ii) *subnet agent* (SA). The MA provides a domain-wide stable access point for an MN. An SA handles the mobility of MNs within a subnet. Similar to HMIP, each MN can get two CoAs - global CoA (GCoA) and local CoA (LCoA). The GCoA specifies the domain to which the MN is currently attached. The LCoA identifies the MN's present subnet. The packets destined to an MN are first received by the HA. The HA tunnels the packets to the MA using the MN's GCoA. The MA first decapsulates the packets, determines the current LCoA of the MN using its internal table, and tunnels them to the LCoA. The encapsulated packets are received by the SA. Finally, the SA decapsulates the packets and forwards them to the MN. When the MN moves from one subnet to another inside the same domain, it is assigned a new LCoA. The MN registers the address of the new LCoA with its MA. Till the registration of the new LCoA is complete, the MA forwards all packets for the MN to the old LCoA. This results in packet drops. A fast handoff procedure has been proposed to avoid this packet loss (Misra et al., 2002). It eliminates intra-domain update delay by anticipating the handover in connectivity between the networks and the MNs. The anticipation of MN's movement is based on a link layer trigger which initiates a network layer handoff before the link layer handoff completes. Once the MN senses a handoff, it sends a request to the MA to multicast the packets to its SAs. The MA multicasts incoming packets to each neighboring SAs. Each SA buffers the packets in order to prevent any loss of packets in transit during the handoff. After the MN finishes registration, the new SA transfers all buffered packets to the MN.

**Cellular IP:** Cellular IP (Campbell et al., 2000) is a mobility management protocol that provides access to a Mobile IP-enabled Internet for fast moving MHs. The architecture of Cellular IP is shown in Figure 5. It consists of three major components: (i) cellular IP node or the base station (BS), (ii) cellular IP gateway (GW), and (iii) cellular IP mobile host (MH). A Cellular IP network consists of interconnected BSs. The BSs route IP packets inside the cellular network and communicate with MHs via wireless interface. The GW is a cellular IP node that is connected to a regular IP network by at least one of its interfaces. The BSs periodically emit beacon signals. MHs use these beacon signals to locate the nearest BSs. All IP packets transmitted by an MH are routed from the BS to the GW by hop-by-hop shortest path routing, regardless of the destination address. The BSs maintain route cache. Packets transmitted by the MH create and update entries in BS's cache. An entry maps the MH's IP address to the neighbor from which the packet arrived to the host. The chain of cached mappings referring to an MH constitutes a reverse path for downlink packets for the MH.

To prevent timing out of these mappings, an MH periodically transmits control packets. MHs that are not actively transmitting or receiving data themselves may still remain reachable by maintaining paging caches. MHs listen to the beacons transmitted by BSs and initiate handoff based on signal strength. To perform a handoff, an MH tunes its radio to the new BS and sends a *route update* packet. This creates routing cache mappings on route to the new BS. Handoff latency is the time that elapses between the handoff and the arrival of the first packet through the new route. The mappings associated with the old BS are cleared after the expiry of a timer. Before the timeout, both the old and new downlink routes remain valid and packets are delivered through both the BSs. This feature used in Cellular IP semi-soft handoff algorithms improves handoff performance.

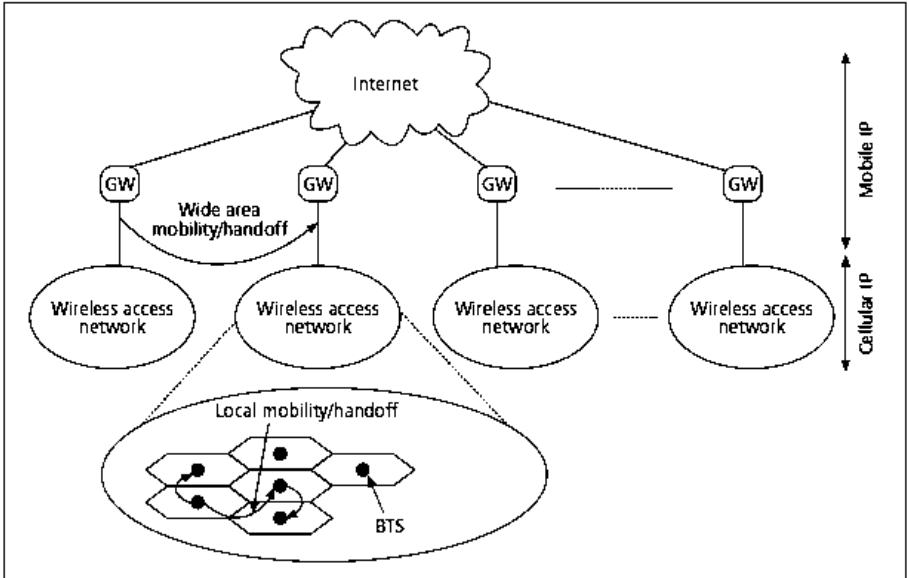

Fig. 5. Architecture of Cellular IP [Source: (Akyildiz et al., 2004)]

**Handoff Aware Wireless Access Internet Infrastructure:** Handoff-Aware Wireless Access Internet Infrastructure (HAWAII) (Ramjee et al., 2002) is a domain-based approach for supporting mobility. The network architecture of HAWAII is shown in Figure 6. Mobility management within a domain is handled by a gateway called a *domain root router* (DRR). Each MH is assumed to have an IP address and a home domain. While moving in its home domain, the MN retains its IP address. The packets destined to the MH reach the DRR based on the subnet address of the domain and are then forwarded to the MH. The paths to MH are established dynamically. When the MH is in a foreign domain, packets for the MH are intercepted by its HA. The HA tunnels the packets to the DRR of the MH. The DRR routes the packets to the MH using the host-based routing entries. If the MH moves across different subnets in the same domain, the route from the DRR to the BS serving the MN is modified, while the other paths remain unchanged. This causes a reduction in signaling message and handoff latency during intra-domain handoff. In traditional Mobile IP, the MH is directly attached either to the HA (i.e. the home domain router) or the FA (i.e. the foreign domain router). Thus, every handoff causes a change in the IP address for the MH, resulting in lack of scalability. HAWAII also supports IP paging. It uses IP multicasting to page idle MHs when packets destined to an MH arrive at the domain root router and no recent routing information is available.

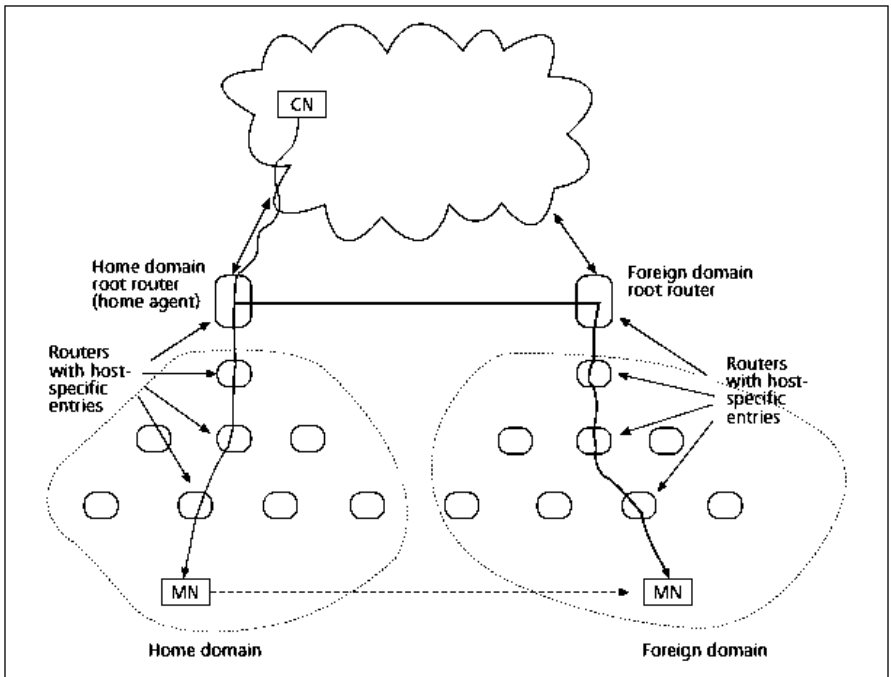

Fig. 6. Architecture of HAWAII Protocol [Source: (Akyildiz et al., 2004)]

**Summary**: Various network layer micro-mobility management schemes have been compared based on their features (Chiussi et al., 2002; Ramjee et al., 1999; Campbell et al., 2002). Each protocol uses the concept of domain root router. In all the protocols, signaling traffic is largely localized in a domain so as to reduce the global signaling traffic overhead. Routing-based schemes utilize the robustness of IP forwarding mechanism. Mobile-specific address lookup tables are maintained by all the mobility agents within a domain. In tunnel-based schemes, registration of the mobile nodes and encapsulation of the IP packets are performed in a local or hierarchical manner. Routing-based schemes avoid tunneling overhead, but suffer from the high cost of propagating host-specific routes in all routers within the domain. Moreover, the root node in routing schemes is a potential single point of failure (Chiussi et al., 2002). Tunnel-based schemes are modular and scalable. However, they introduce more cost and delays (Campbell et al., 2002).

## 4. Link Layer Mobility Management Mechanisms

Link layer mobility management mechanisms deal with issues related to inter-system roaming between heterogeneous access networks with different radio technologies and network management protocols. Two important considerations for designing inter-system roaming standards are: (i) the protocols for air interface and (ii) the mobile application part (MAP). In situations where a mobile node enters one wireless access network from another that support the same air interface protocols and MAP, the services are seamlessly migrated.

However, when the MAPs are different for the two networks, additional network entities need to be placed and signaling traffic are to be transmitted for inter-working. Since each network has its own mobility management protocols, the new inter-working entities should not replace existing systems. Rather, the entities should coexist and inter-work.

### 4.1 Location management protocols

For next-generation heterogeneous wireless networks, the inter-working and inter-operating function is suggested to accommodate roaming between dissimilar networks (Pandya et al., 1997). For existing practical systems, several solutions are proposed for some specific pairs of inter-working systems. In these schemes, the inter-operating function is implemented in either some additional inter-working unit with the help of dual-mode handsets (Phillips & Namee, 1998), or a dual-mode home location register (HLR) (Garg & Wilkes, 1996) to take care of the transformation of signaling formats, authentication, and retrieval of user profiles. Recent research efforts attempt to design general location management mechanisms for the integration and inter-working of heterogeneous networks. The research activities can be grouped into two categories: location management for adjacent dissimilar systems with partially overlapping coverage at the boundaries (Akyildiz & Wang, 2002; Wang & Akyildiz, 2001; ETSI, 2002) and location management in multi-tier systems where service areas of heterogeneous networks are fully overlapped (Lin & Chlamtac, 1996). All these solutions propose additional entities that take care of inter-working issues.

**Location Management for Adjacent Networks:** Researchers have addressed the issues of location management in two adjacent networks with overlapping areas (Akyildiz & Wang, 2002; Wang & Akyildiz, 2001; ETSI, 2002). Some of the protocols are discussed briefly.

**Gateway Location Register Protocol:** To enable inter-system roaming, a new level has been introduced in the hierarchy of location management entities for UMTS/ IMT-2000 networks. The new level consists of a *gateway location register* (GLR) (ETSI, 2002). The GLR is a gateway that enables inter-working between two networks by suitably converting signaling and data formats. It is located between the *visitor location register* (VLR) and the *serving GPRS support node* (SGSN) and the *home location register* (HLR). When a subscriber roams, the GLR plays the role of the HLR toward the VLR and SGSN in a *visited public land mobile network* (VPLMN), and the role of the VLR and SGSN to the HLR in a *home public land mobile network* (HPLMN). The GLR protocol assists the operators in lowering costs and optimizing roaming traffic. However, the protocol is not designed for ongoing call connection during inter-system roaming (Wang & Akyildiz, 2001). The incoming calls are routed to the home network even when the MN is roaming. This makes the protocol inefficient.

**Boundary Location Register Protocol:** In (Akyildiz & Wang, 2002), a location management mechanism has been proposed for heterogeneous network environment. It involves a mechanism for inter-system location updates and paging. Inter-system location update is implemented by using the concept of a *boundary location area* (BLA) existing at the boundary between two systems - *X* and *Y* in Figure 7. The BLA is controlled by a *boundary interworking unit* (BIU), which is connected to the *mobile switching centers* (MSCs) in both the systems. The BIU queries the user's service information, converts the message formats, checks the compatibility of the air interfaces and performs authentication of mobile users. When an MN is inside its BLA, it sends a location registration request to the new system. A distance-based location update mechanism reports MN's location when its distance from the boundary is less than a pre-defined threshold. An entity called a *boundary location register* (BLR) is used for inter-system paging. The BLR maintains in its cache the location information of the MN and its roaming information when it crosses an intersystem boundary. During the inter-system paging process, only one system (*X* or *Y*) is searched. The associated MAP protocol is designed for mobile nodes with ongoing connections during inter-system roaming (Wang & Akyildiz, 2001). Instead of performing location registration after a mobile node arrives at the new system, the BLR protocol enables the node to update its location and user information actively before it enters the new system. In this way, the incoming calls to the MN during its inter-system roaming are delivered to the node.

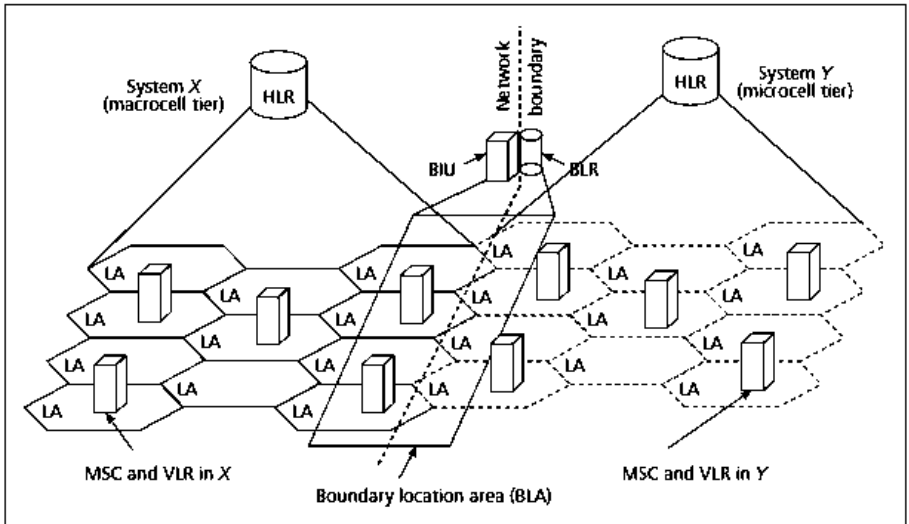

Fig. 7. The Boundary Location Register Protocol [Source: (Akyildiz et al. 2004)]

**Location Management in Heterogeneous Networks:** An MN is reachable via multiple networks when their service areas are fully overlapped. Since heterogeneous networks use different signaling formats, authentication procedures, and registration messages, it is difficult to merge heterogeneous HLRs into a single HLR. A multi-tier HLR (MHLR) is proposed in (Lin & Chalmtac, 1996), where a *tier manager* is connected to all the HLRs. Two types of location registration are possible: (i) single registration (SR) and (ii) multiple registrations (MR). Under SR scheme, an MN associates with the lowest tier of the MHLR, and receives services at low cost and high bandwidth. Under MR method, the MN registers on multiple tiers simultaneously. The individual tiers perform their own roaming management. The tier manager keeps track of the currently visited high-tier and low-tier VLRs of the MN. It has been found that MR scheme involves less signaling overhead (Lin & Chlamtac, 1996). However, since the current tier of the MN is not known to the MHLR, it incurs a high loss when a wrong tier is selected.

**Summary:** To summarize, all link layer-based mobility management schemes require additional inter-working entities for enabling information exchange between different systems. These inter-working entities are different depending on the systems, e.g., the GLR/BLR for inter-system location management, the MHLR for a multi-tier PCS system, and the gateways in the integrated UMTS/WLAN system. The interworking entities perform the following functions: (i) format translation of the signaling messages and data packets and address translation between networks, (ii) retrieval of user profile from the home network, (iii) acting as a gateway for signal transmission and route setup, (iv) recording of mobility-related information during inter-system roaming, (v) negotiating QoS when an MN enters a new network, and (vi) performing authentication during inter-system movement. Different approaches for mobility management at the link layer address the following issues: (i) the location where the inter-working entities are put, (ii) the degree of coupling (loose or tight) of the entities, (iii) the timing of location registration and handoff initiation, (iv) the way location and handoff management is performed.

## 5. Handoff Management Protocols

Handoff or handover is a process by which an MN moves from one point of network attachment to another. Handovers can be classified as either homogeneous or heterogeneous. A heterogeneous handover occurs when an MN either moves between networks with different access technologies, or between different domains. As the diversity of available networks increases, it is important that mobility technologies become agnostic to link layer technologies, and can operate in an optimized and secure fashion without incurring unreasonable delay and complexity (Dutta et al., 2008). Supporting handovers across heterogeneous access networks, such as IEEE 802.11 (Wi-Fi), global system for mobile communications (GSM), code-division multiple access (CDMA), and worldwide interoperability for microwave access (WiMAX) is a challenge, as each has different quality of service (QoS), security, and bandwidth characteristics. Similarly, movement between different administrative domains poses a challenge since MNs need to perform access authentication and authorization in the new domain. Thus, it is desirable to devise a mobility optimization technique that can reduce these delays and is not tightly coupled to a specific mobility protocol. In this section, we describe different types of handovers and investigate the components that contribute to a handover delay. Some inter-technology and media-independent handover frameworks are then described.

### 5.1 Taxonomy of handoff mechanisms
Different types of handovers may be classified based on three parameters as follows: (i) subnets, (ii) administrative domains, and (iii) access technologies (Dutta et al., 2008).
*Inter-technology*: this type of handover is possible with an MN that is equipped with multiple interfaces supporting different technologies. An inter-technology handover occurs when the two points of attachment use different access technologies. During the handoff, the MN may move out of the range of one network (e.g., Wi-Fi) into that of a different one (e.g., CDMA). This is also known as *vertical handover*.
*Intra-technology*: this type of handoff occurs when an MN moves between points of attachments supporting the same access technology, such as between two Wi-Fi access points. An intra-technology handover may happen due to intra-subnet or inter-subnet movement and thus may involve the layer 3 trigger.
*Inter-domain*: when the points of attachment of an MN belong to different domains, this type of handoff takes place. A domain is defined as a set of network resources managed by a single administrative entity that authenticates and authorizes access for the MNs. An administrative entity may be a service provider or an enterprise. An inter-domain handover possibly involves an inter-subnet handover also.

*Intra-domain*: handovers of this type occurs when the movement of an MN is confined within an administrative domain. Intra-domain movement may also involve intra-subnet, inter-subnet, intra-technology, and/or inter-technology handovers as well.

*Inter-subnet*: an inter-subnet handover occurs when the two points of attachment belong to different subnets. The MN acquires a new IP address and possibly undergoes a new security procedure. A handover of this type may occur along with either an inter- or an intra-domain handover and also with either an inter- or an intra-technology handover.

*Intra-subnet*: an intra-subnet handover occurs when the two points of attachment belong to the same subnet. This is typically a link layer handover between two access points in a WLAN networks, or between different cell sectors in cellular networks. It is administered by the radio network and requires no additional authentication and security procedures.

### 5.2 Delays in handoff

All the layers in the communication protocol stack contribute to the delay in a handoff.

*Link layer delay*: depending on the access technology, an MN may go through several steps with each step adding its contribution to the overall delay before a new link is established. For example, a Wi-Fi link goes through the process of scanning, authentication, and association before being attached to a new access point. For intra-subnet handovers, where network layer configurations are necessary, link layer contributes the maximum to the overall delay.

*Network layer delay*: after completion of the link layer procedures, it may be necessary to initiate a network layer transition. A network layer transition may involve steps such as: acquiring a new IP address, detecting a duplicate address, address resolution protocol (ARP) update, and subnet-level authentication.

*Application layer delay*: the delay of this type is due to reestablishment and modification of the application layer properties such as IP address while using *session initiation protocol* (SIP). The authentication and authorization procedure such as *extensible authentication protocol* (EAP) includes several round-trip messages between the MN and the *authentication authorization and accounting* (AAA) server causing delay in handoff.

### 5.3 Research work on handoff mechanisms

This section presents some of the existing handoff mechanisms proposed in the literature.

In (Hasswa et al., 2005), a vertical handoff decision function is proposed for roaming across heterogeneous wireless networks. An optimization scheme for vertical hand off has been proposed in (Zhu & McNair, 2004). In (Park et al., 2003), a seamless vertical handoff scheme is proposed between a WLAN and a CDMA 2000-based cellular network. A vertical handoff scheme between a UMTS and a WLAN network is proposed in (Zhang et al., 2003). A *connection manager* detects the changes in wireless networks and makes the handoff decision. When the MN moves from the UMTS to the WLAN network, the objective of the handoff is to have better QoS because of the higher bandwidth of WLAN. However, in case of handoff from WLAN to UMTS, the handoff is initiated just before the connection to WLAN breaks.

In (Efthymiou et al., 1998), a protocol for *inter-segment handover* (ISHO) is proposed in an integrated space/terrestrial UMTS environment. A backward mobile-assisted handover incorporating signalling diversity is chosen as the most appropriate handover scheme. Based on the *generic radio-access network* (GRAN) concept and by using a satellite-UMTS network architecture and functional model, the derivation of an ISHO protocol is presented.
In (McNair et al., 2000), a handoff technique is introduced that supports mobility between networks with different handover protocols. Three types of handoffs are presented: (i) *network-controlled handoff* (NCHO), (ii) *mobile-assisted handoff* (MAHO), and (iii) *mobile-controlled handoff* (MCHO). Under NCHO or MAHO, the network generates a new connection, finds new resources for the handoff and performs any additional routing operations. For MCHO, the MN finds the new resources and the networks approves.
In (Stemm & Katz, 1998), a vertical handoff scheme is designed for wireless overlay networks, where heterogeneous networks in a hierarchical structure have fully overlapping service areas. The BSs send out periodic beacons similar to Mobile IP FA advertisements. The MN listens to these packets and decides which BS would forward packets, which BS should buffer packets for a handoff, and which BS should belong to the multicast group.
In (Buddhikot et al., 2003), the issues of integration of WLAN and 3G networks have been addressed to offer seamless connectivity. Two approaches have been identified: (i) a tightly-coupled approach and (ii) a loosely-coupled approach. In the tightly-coupled approach, the gateway of 802.11 network appears to the upstream 3G core as either a *packet control function* (PCF), in case of a CDMA2000 core network, or as a *serving GPRS service node* (SGSN), in case of a UMTS network. The 802.11 gateway hides the details of the 802.11 network to the 3G core, and implements all the protocols required in a 3G access network. In the loosely-coupled scheme, the same 802.11 gateway is used. However, the gateway connects to the Internet and does not have any direct link to the 3G network elements such as *packet data service nodes* (PDSNs), *gateway GPRS service nodes* (GGSNs) or 3G core network switches. In this case, the data paths in 802.11 and 3G networks are different. The high speed 802.11 traffic is never injected into the 3G network but the end user still achieves seamless access.

### 5.4 Cross-layer handoff mechanisms

The cross-layer protocols for mobility management are mainly applied for handoff. Most of these mechanisms use link layer information to make an efficient network layer handoff. The utilization of link layer information reduces the delay in movement detection of the MN so that the overall handoff delay is minimized.
In (Yokota et al., 2002), a low-latency handoff algorithm for a WLAN has been proposed that uses access points and a dedicated *medium access control* (MAC) bridge. A seamless handoff architecture for Mobile IP, called S-MIP is presented in (Hsieh et al., 2003) that combines a location tracking scheme with the HMIP handoff. A vertical handoff mechanism between IEEE 802.11 (WLAN) and IEEE 802.16e (Mobile WiMAX) networks in a wireless mesh backbone is proposed in (Zhang, 2008). In (Dutta et al., 2008), a media-independent pre-authentication scheme has been proposed. These four handoff schemes are discussed below.
**Link Layer-Assisted Fast Handoff over WLAN:** In the Mobile IP protocol, the MN movement can be detected from advertisements of the FAs that differ from the previously received advertisement. The new CoA is registered with the HA. However, data packets are not forwarded to the new FA before the registration is complete. This interruption may degrade the QoS especially in real-time applications. To tackle this issue, a handoff mechanism is proposed in which APs in a WLAN and a dedicated MAC bridge are jointly used to eliminate packet loss (Yokota et al., 2002). The authors have noted that the delay in Mobile IP handoff is contributed by two elements: (i) the delay in movement detection of the MN, and (ii) delay due to signaling for registration. The proposed mechanism reduces the movement detection delay. It has two parts: (i) handoff for the forward direction (i.e. mobile-terminated data) and (ii) handoff for the reverse direction (i.e. mobile-originated data). The APs in the WLAN have the capability to notify the MAC address of an MN that moves into their coverage areas. The MAC bridge is configured in a way that it sends only those MAC frames whose destination addresses are registered in the filtering database (DB).

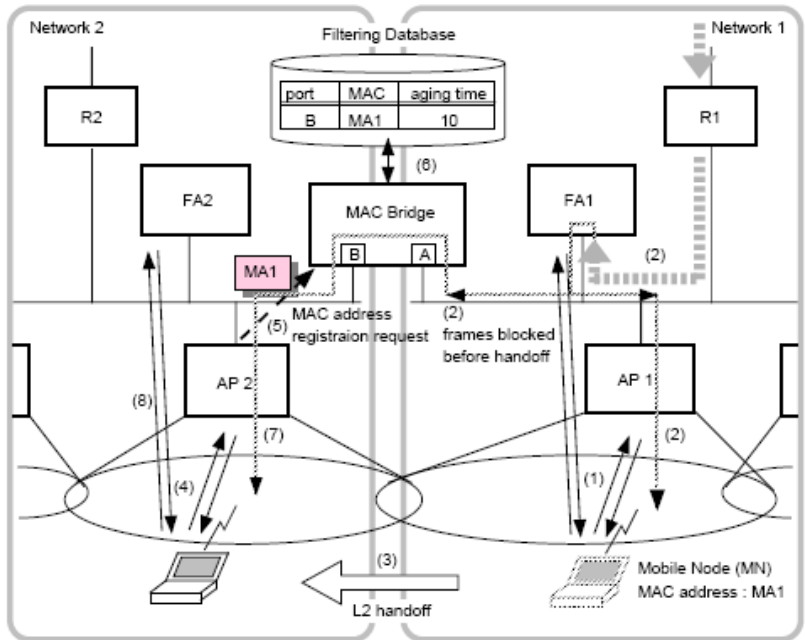

Fig. 8. Handoff Scenario in Forward Direction [Source: (Yokota et al., 2002)]

The handoff in the forward direction happens as follows. In Figure 8, the MN establishes an association with an access point- AP1, and registers the CoA with HA. The packets destined to the MN are encapsulated by the HA and tunneled to FA1- the FA of the MN. FA1 decapsulates the packets and sends them directly to the MN. When the signal strength of the channel of communication between AP1 and the MN falls below a threshold, MN attempts to find a new AP. The MN establishes association with a new AP- AP2. AP2 places the MAC address of the MN in a MAC *address registration request message* and broadcasts it on the local segment. The MAC bridge receives the address registration request. It then makes an entry of the MAC address contained in the message and the port on which the message was received into the filtering DB. When the MAC bridge receives a MAC frame on a port, it refers to the filtering DB to see if the destination MAC address is registered. If the address is registered, the MAC bridge sends it out to the corresponding port. Packets from FA1 are thus bridged from port *A* to the port *B* of the MAC bridge, and delivered to the Network 2, to which the MN is now connected. The MN detects its movement as it receives new *agent advertisements* from FA2 and registers the new CoA with the HA. When the registration is complete, packets destined for the MN are tunneled to FA2 and delivered to the MN. Since no packets are bridged from that time onward, the entry for the MN in the filtering DB must be removed upon expiration of its aging time. Thus, the MN receives packets even before Mobile IP registration is over.

If the MAC bridge relays only those frames whose source MAC addresses are registered in the filtering DB to the network to which MN was previously attached, then it can reduce transmission interruption in the reverse direction as well. However, the transmission interruption in the reverse direction is possible if the MAC bridge has only two ports. The MAC bridge with two ports checks the source MAC address of an incoming frame from one port with the filtering DB, and transfers it to the other port. However, if the MAC bridge has more than two ports, the direction in which the frame should be transferred will depend on the speed of the MN and how fast the Mobile IP registration process completes. By taking into account that the next hop of a frame sent by the MN is always the default router of the network where the MN has been registered, the authors have proposed a fast handoff method in the reverse direction by registering the MAC address of the default router in the filtering DB. The algorithm exploits the Mobile IP agent advertisement message which are periodically broadcasted by the FAs and received by the MN. The scheme has been evaluated in an actual network environment to measure the time required for forward and reverse handoffs on UDP and TCP traffic. The latency due to Mobile IP handoff has been found to be equal to that of a link layer handoff (Yokota et al., 2002).

**Seamless Handoff Architecture for Mobile IP:** Seamless Handoff Architecture for Mobile IP (S-MIP) is an architecture which minimizes the handoff latency in a large indoor environment (Hsieh et al., 2003). The architecture of S-MIP is depicted in Figure 9. It is an extension of the HMIP architecture with an additional entity called *decision engine* (DE). The DE is identical to MAP in HMIP, and makes the handoff decision for its network domain. The MAP separates the mobility type into micro-mobility and macro-mobility. The *new access router* (nAR) and the *old access router* (oAR) retain the same functionality and meaning as in HMIP. Through periodic feedback information from the ARs, the DE maintains a global view of the connection state of any MN in its network domain. DE also tracks the movement patterns of all MNs in its domain using the signal strength information received from the link layer and the IDs of the ARs.

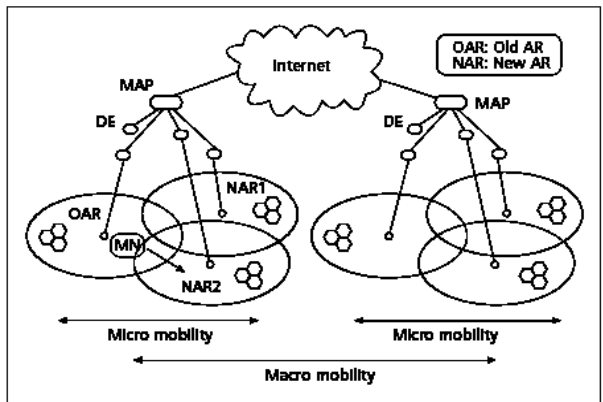

Fig. 9. Architecture of the S-MIP Scheme [Source: (Hsieh et al., 2003)]

In HMIP and fast handoff mechanisms, the packet loss occurs either within the MAP and the ARs (segment packet loss), or between the last ARs and the MN (edge packet loss). While edge packet losses occur due to the mobility of an MN and transmission errors, the segment-packet loss is due to the non-deterministic nature of handoffs and the resulting switching of the data stream at the MAP after the receipt of the MAP *binding update*. The design of S-MIP minimizes the edge packet and segment packet losses. Edge packet loss is minimized by keeping the anchor point for the forwarding mechanism as close to the MN as possible. Hence it is located at the AR that bridges the wireless network and the wired network. Segment packet loss is minimized by using a newly developed *synchronized packet simulcast* (SPS) scheme and a hybrid handoff mechanism. The SPS simulcasts packets to the current network where the MN is attached to and to the potential access network that the MN is asked to switch onto. The hybrid handoff strategy is MN-initiated, but network determined. The decision as to which access network to handoff is formulated from the movement tracking mechanism which is based on a synchronized feedback. The authors have provided a combination of simulation results and mathematical analysis to argue that S-MIP is capable of providing zero-packet loss handoff with latency similar to that of a link layer delay in a WLAN environment.

**A Vertical Handoff Scheme between WLAN and Mobile WiMAX Networks:** In (Zhang, 2008), a vertical handoff scheme has been proposed between 802.11(WLAN) and 802.16e (Mobile WiMAX) networks. The framework has been discussed with a *wireless mesh network* (WMN) that provides high speed, scalable and ubiquitous wireless Internet services. A *wireless mesh router* (WMR) is a gateway that has routing capabilities to support mesh networking. Each WMR is assumed to have 802.11e functions, 802.16e BS functions with *point-to-multi-point mode* (PMP), routing capabilities, and 802.16e subscriber station (SS) functions with mesh mode. The MNs can connect only via mesh routers to access the Internet using two types of links: the IEEE 802.11e and IEEE 802.16e links. The IEEE 802.16e links between MNs and mesh routers operate in the PMP mode, while the IEEE 802.16e links among neighboring mesh routers operate in the mesh mode. Figure 10 illustrates the system. The links between the WMRs are 802.16e mesh links. The WMR which is connected to the Internet with wired line is called *mesh gateway* (MGW). The MNs with dual network interfaces can connect to the Internet through the WMRs by an 802.11e link or 802.16e link. The WMRs which are connected directly or indirectly with one MGW form a domain or subnet. The MNs connect to the WMRs using 802.11e link for high data rate and small coverage area and 802.16e links for higher data rate and large coverage.

An MN initially sets up a connection with a WMR. The WMR forwards the IP packets from the MN to the MGW through one or more WMRs. The MGW transmits the IP packets to the CN in the Internet. IP packets from the CN are routed through the reverse route to the CN. The CN may be located in the same domain as the MN. In this case, the WMRs forward the IP packets for them. While an MN is inside the area doubly covered by the WLAN and WiMAX, a proper vertical handoff is needed if the WLAN network is congested or if the MN is roaming across the edge of the WLAN coverage. The author has proposed a vertical handoff scheme for this scenario. The algorithm has four steps: (i) new network interface scanning, (ii) new access router discovery, (iii) new network entry, and (iv) routing information updating. After completion of these stages, the MN can transmit or receive information data packets through the new network interface.

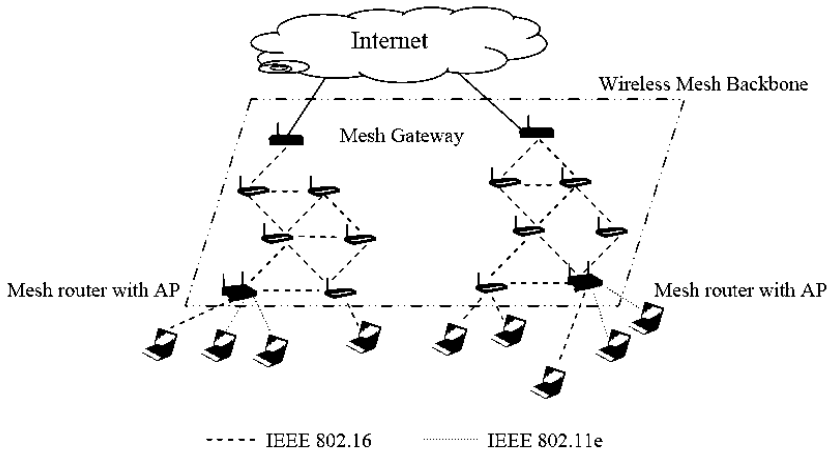

Fig. 10. Architecture of the Wireless Mesh Network [Source: (Zhang, 2008)]

In Figure 11, two domains are served by service providers *A* and *B* respectively. The WMRs in the same and different domains are called *intra-mesh routers* and *inter-mesh routers* respectively. If the CN is in the same domain as the MN, the IP packets are routed through the intra-mesh routers only. When the MN moves to another domain, the packets from the CN are routed via the HA. Four scenarios are considered for MN mobility.

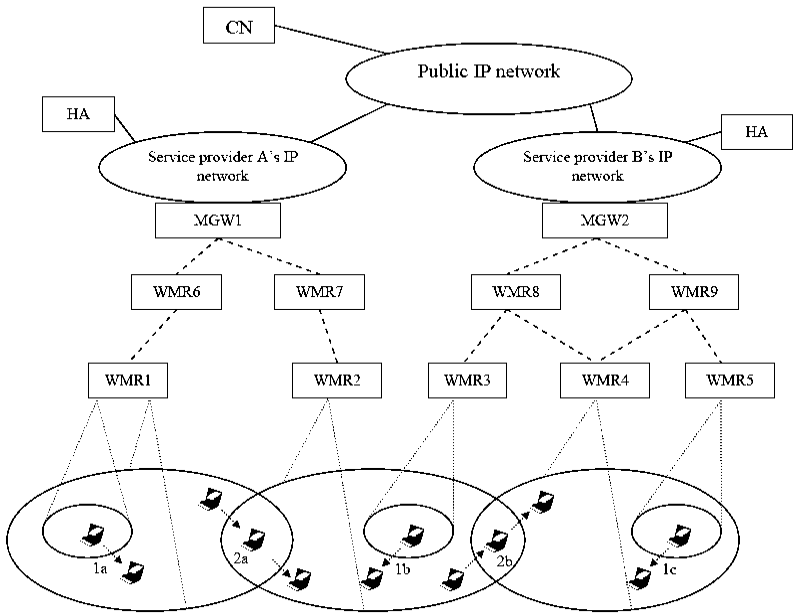

Fig. 11. Macro-Mobility and Micro-Mobility Scenario [Source: (Zhang, 2008)]

*Scenario 1*: the MN is connected to the WLAN. It moves out of WLAN and connects to the WiMAX. The movements 1a, 1b, 1c depict this situation. The WMR does not change, only the medium access interface changes in case of 1a. The handoff occurs between intra-mesh routers in 1c and between inter-mesh routers in case 1b.

*Scenario 2*: the MN is currently connected to the WiMAX. It moves into the WLAN and either connects to the WLAN or continues with the WiMAX connection depending on the network conditions, user preference, or application QoS requirements.

*Scenario 3*: the MN is located in the double-coverage area (i.e. area covered by WLAN and WiMAX) and is currently stationary. If the WLAN is congested, the MN can switch to the WiMAX if it can provide more bandwidth for the MN to transmit its data packets.

*Scenario 4*: A horizontal handoff occurs when the MN moves in 2a and 2b. In (Kim et al., 2005), a scheme called *last packet marking* (LPM) has been proposed for case 2a. The MIPSHOP (Mobility for IP: Performance, Signaling and Handoff Optimization) working group of the Internet Engineering Task Force (IETF) has developed Mobile IPv6 fast handoff over 802.16e networks for case 2b (Jang et al., 2008).

**Media Independent Pre-Authentication for Secure Inter-Domain Handover:** A *media-independent pre-authentication* (MPA) scheme has been proposed in (Dutta et al., 2008). It is a mobile-assisted, secure handover optimization scheme that works over any link layer and with any mobility management protocol. With MPA, an MN securely obtains an IP address and other configuration parameters for a *candidate target network* (CTN) - the network to which the mobile node is being handed off. The MN is also able to send and receive IP packets using the IP address before it attaches to the CTN. In this way, the MN completes the binding update and use the new CoA before performing a handover at the link layer.

MPA provides four basic procedures that optimize handover for an MN. The serving network is the network that currently serves the MN. The first procedure - *pre-authentication* establishes a security association with the CTN to secure subsequent protocol signaling. The second procedure - *pre-configuration* securely executes a configuration protocol to obtain an IP address and other parameters from the CTN. The third procedure executes a *tunnel management protocol* that establishes a *proactive handover tunnel* (PHT) between the MN and an access router in the CTN over which binding updates as well as data packets, can travel. Finally, the fourth procedure deletes the PHT before attaching to the CTN and reassigns the inner address of the deleted tunnel to its physical interface after the MN attaches to the target network. The final two procedures are collectively referred to as *secure proactive handover*. Through the third procedure the MN completes higher-layer handover before starting link layer handover. This means that the MN is able to perform all the higher-layer configuration and authentication procedures before link layer connectivity to the CTN is established. This can significantly reduce the handover delays.

**Summary:** As a macro-mobility management protocol, Mobile IP is simple, but it has several shortcomings such as triangular routing, high-global signaling load, and high handoff latency. Although, the route optimization mechanism eliminates triangular routing, the high handoff latency still remains. The micro-mobility management mechanisms are not suitable for inter-domain mobility. Most of these solutions assume one domain to be one wireless access network or under one administrative domain. Although IDMP (Misra et al., 2002) defines a domain based on geographic proximity where one domain consists of networks with different access technologies in a particular geographic region, there is no procedure specified for inter-system authentication, format transformation, and so on. In a heterogeneous environment where users have freedom to move between different domains, the global signaling load and corresponding handoff delay will increase significantly, adversely affecting the network performance. The S-MIP approach (Hsieh et al., 2003) demonstrates that along with the hierarchical architecture and procedures for fast handoff, the link layer information used to determine the mobility pattern of the MHs can greatly improve intra-domain handoff performance. However, the protocol cannot be extended to support mobility between different domains, because the coverage area of one domain might be completely covered by another domain in the hierarchical heterogeneous environment; for example, a WLAN domain is mostly covered completely by the overlaying 2G/3G network.

## 6. IEEE 802.21- Media Independent Handover Services

A novel solution that ensures interoperability between several types of wireless access network is given by the developing IEEE 802.21 standard (Eastwood et al., 2008). The work on the standard began in 2004 and is expected to be finalized around 2010. The IEEE 802.21 is focused on handover facilitation between different wireless networks in heterogeneous environments. The standard names this type of vertical handover as Media Independent Handover (MIH). In MIH, the handover procedures can use the information gathered from both the mobile terminals and the network infrastructure. At the same time, several factors may determine the handover decision, e.g., service continuity, application class and QoS, negotiation of QoS, security, power management, handover policy etc. IEEE 802.21 facilitates, speeds, and thereby increases the success rate of inter-technology handover decision making and other pre-execution processes. These processes include inter-technology candidate network discovery, target network selection, target network preparation, and handover execution timing and initiation. IEEE 802.21 defines three services to facilitate inter-technology handovers: (i) *media independent information service* (MIIS), (ii) *media independent command service* (MICS), and (iii) *media independent event service* (MIES). MIIS provides information about the neighboring networks, their capabilities and available services. MICS allows effective management and control of different link interfaces on multimodal device and enables both mobile- and network-initiated handovers. It supports querying of target networks about the status of the rapidly changing resources. Some MICS commands are part of the signaling between inter-*radio access technology* (RAT) gateways. MIES provides events triggered by changes in the link characteristics and status.
This interface provides service primitives to the upper layers that are independent of the access technology.
One of the most important aspects of MIH is the fact that it allows for *network controlled handovers* and *user controlled handovers*. The advantages of the network controlled handover lies in the lower user battery consumption since the monitoring of various network conditions is done by the networks themselves. However, it incurs a huge signaling overhead and a high processing load in the network elements. In user controlled handover, the user collects necessary data and initiates the appropriate actions. The disadvantage of this approach is the high battery power consumption.

**6.1 Mobility using IEEE 802.21 in a heterogeneous IMT-advanced (4G) network**
The telecommunication industry is defining a new generation of mobile wireless technologies, called *fourth generation* (4G). In this regard, the International Telecommunications Union- Radio Standardization Sector (ITU-R) has defined the concept of IMT-Advanced that targets peak data rates of about 100 Mb/s for highly mobile access (at speeds of up to 250 km/hr), and 1 Gb/s for low mobility (pedestrian speeds or fixed) access. The IEEE is developing extensions to both IEEE 802.11 and 802.16 to meet IMT- Advanced requirements. The evolving standard of IEEE 802.16m aims to achieve a data rate of 100 Mb/s in a highly mobile (25 km/hr) scenario. These data rate and mobility capabilities make 802.16m a candidate for the high mobility portion of the IMT-Advanced standard requirements. Another working group of IEEE 802.11n is working towards designing a *very high throughput* (VHT) radio capable of data rates up to 1 Gb/s at stationary or pedestrian speeds. Together, 802.16m and 802.11n will satisfy both the low-mobility and fully mobile user velocity vs. data rate requirements for IMT-Advanced systems. If IEEE proposes a combination of 802.11m and 802.11n for IMT-Advanced standard, an interworking mechanism must be designed for tying up these two systems. In (Eastwood et al., 2008), the authors have proposed a mobility management approach in 4G using IEEE 802.21 Media Independent Handover (MIH) services.

## 7. Security in Handoff Procedures

Whenever an MN connects to a point of network access, it establishes a security context with the service provider. During the handover process, some or all the network entities involved in the security mechanism may change. Thus the current security context changes as well. The MN and the network have to ensure that they still communicate with each other and they agree upon the keys to protect their communication.
However, during handovers in networks like GSM/GPRS and UMTS no authentication is used. This makes the handover procedures vulnerable to a hijacking attack. An attacker can masquerade as an authentic mobile station (MS) just by sending message at the right frequency and time slot during handover. As long as the attacker does not know the encryption and/or integrity keys currently being used, he cannot insert valid traffic into the channel. However, if an attacker can gain access to the key(s) (e.g. because of a missing protection on the backbone network), he can impersonate the MS. In fact, in GSM/GPRS, UMTS and WLAN networks, no standard protection mechanism in the backbone network has been specified. Many GSM operators do not protect the radio link between their fixed networks and the BSs. In UMTS, during a handover, the keys used to protect the traffic between the MS and the previous BS are reused in communication with the next BS. While the keys are being transmitted, they can be intercepted by an adversary, if the wireless link is not protected.
Usually an authentication process happens before location updates and call setups. The same mechanisms cannot however, be applied in establishing connection during a handover process because of the stringent time constraint. In GSM, for example, the time between the handover command and the handover complete or handover failure message is restricted to 0.5- 1.5 s. The generation of an authentication response, however, takes about 0.5 s at the MS side. Thus an authentication overhead will cause connection disruption.

As we have seen earlier in this chapter, efficient cell prediction mechanisms can reduce the signaling overhead between the MS and the old BS. The free time slots may be used to forward authentication traffic between the MS, the old BS and the new BS. The MS can pre-compute an authentication challenge and the encryption and integrity protection keys before the actual change of channel. When the MS and the new BS establish connection, the MS sends the pre-computed authentication response for the new BS to check. If the checking yields positive results, a *handover complete* message is sent and the old BS releases its resources. Otherwise, a *handover failure* happens and the MS falls back to the old channel.

## 8. Some Open Issues in Mobility and Handover Management

Future wireless networks will be based on all-IP framework and heterogeneous access technologies. Design of efficient mobility management mechanisms will be playing ever important role in providing seamless services. Following issues will play dominant roles.
**QoS issues** – next-generation all-IP wireless networks will have to provide guaranteed QoS to mobile terminals. QoS provisioning in a heterogeneous wireless and mobile networks will bring in new problems to mobility management, such as location management for efficient access and timely service delivery, QoS negotiation during inter-system handoff, etc.
**User terminals** – the design of a single user terminal that is able to autonomously operate in different heterogeneous access networks will be another important factor. This terminal will have to exploit various surrounding information (e.g., communication with localization systems, cross-layering with network entities etc.) in order to provide richer user services (e.g. location/situation/context–aware multimedia services). This will also put strong emphasis on the concept of cognitive radio and cognitive algorithms for terminal re-configurability.
**Location and handoff management in wireless overlay networks** – future wireless networks will be inherently hierarchical where access networks have different coverage areas. Mobility management in wireless overlay networks will be a very important issue.
**Mobile services** – sophisticated 4G service discovery mechanisms will combine the location/situation information and context-awareness in order to deliver users' services in a best possible manner. Additionally, future mobile services will require more complex personal and session mobility management to provision personalized services through different personalized operating environments to a single user terminal address. Whether SIP should be the core 4G protocol, and whether the service delivering framework be the network layer-based or application layer-based is still an open question.
**Cross-Layer optimization** – design of efficient cross-layer-based approaches will play a key role is developing new mobility management schemes.
**Other issues** – fault-tolerance, availability of network services, enhanced security, intelligent packet and call routing, intelligent gateway discovery and selection procedures and design of a unified protocol stack and vertical protocol integration mechanisms are some of the other important issues in next-generation heterogeneous networks.

## 9. Conclusion

In this chapter, a comprehensive discussion has been made on mobility management in next-generation wireless networks. Issues in location registration and handoff management have been identified and several existing mechanisms have been presented. Since global roaming will be an increasing trend in future, attention has been paid on mechanisms which are applicable in heterogeneous networks. Media Independent Handover Services of IEEE 802.21 standard as an enabler for handover has also been presented. Security and authentication issues in next-generation heterogeneous networks are discussed briefly. Finally, the chapter concludes by highlighting some open areas of research in mobility management.